# Elucidating The Mechanism of Large Phosphate Molecule Intercalation Through Graphene Heterointerfaces

Jiayun Liang[1], Ke Ma[1], Xiao Zhao[1,2], Guanyu Lu[3], Jake V. Riffle[4], Carmen Andrei[5], Chengye Dong[6], Turker Furkan[7], Siavash Rajabpour[7], Rajiv Ramanujam Prabhakar[8], Joshua A. Robinson[6,7], Magdaleno R. Vasquez Jr.[9] Quang Thang Trinh[10], Joel W. Ager[1,2], Miquel Salmeron[1,2], Shaul Aloni[11], Joshua D. Caldwell[3], Shawna M. Hollen[4], Hans A. Bechtel[12], Nabil Bassim[5,13], Matthew P. Sherburne[1], Zakaria Y. Al Balushi[1,2,†]

1. Department of Materials Science and Engineering, University of California, Berkeley, Berkeley, CA 94720, USA.
2. Materials Sciences Division, Lawrence Berkeley National Laboratory, Berkeley, CA 94720, USA.
3. Department of Mechanical Engineering, Vanderbilt University, Nashville, TN 37235, USA.
4. Department of Physics and Astronomy, University of New Hampshire, Durham, NH 03824, USA.
5. Canadian Centre for Electron Microscopy, McMaster University, Hamilton ON L8S 4L8, Canada.
6. 2D Crystal Consortium, The Pennsylvania State University, University Park, PA 16802, USA.
7. Department of Materials Science and Engineering, The Pennsylvania State University, University Park, PA 16802, USA.
8. Chemical Sciences Division, Lawrence Berkeley National Laboratory, Berkeley, CA 94720, USA.
9. Department of Mining, Metallurgy, and Materials Engineering, University of the Philippines, Diliman, Quezon City 1101, Philippines
10. Queensland Micro- and Nanotechnology Centre, Griffith University, Brisbane, 4111 Australia
11. The Molecular Foundry, Lawrence Berkeley National Laboratory, Berkeley, CA 94720, USA.
12. Advanced Light Source, Lawrence Berkeley National Laboratory, Berkeley, CA 94720, USA.
13. Department of Materials Science and Engineering, McMaster University, Hamilton ON L8S 4L8, Canada.

†To whom correspondence should be addressed. e-mail: albalushi@berkeley.edu

## Abstract

Intercalation is a process of inserting chemical species into the heterointerfaces of two-dimensional (2D) layered materials. While much research has focused on intercalating metals and small gas molecules into graphene, the intercalation of larger molecules through the basal plane of graphene remains highly unexplored. In this work, we present a new mechanism for intercalating large molecules through monolayer graphene to form confined oxide materials at the graphene-substrate heterointerface. We investigate the intercalation of phosphorus pentoxide ($P_2O_5$) molecules directly from the vapor phase and confirm the formation of confined $P_2O_5$ at the graphene heterointerface using various techniques. Density functional theory (DFT) corroborate the experimental results and reveal the intercalation mechanism, whereby $P_2O_5$ dissociates into small fragments catalyzed by defects in the graphene that then permeates through lattice defects and reacts at the heterointerface to form $P_2O_5$. This process can also be used to form new confined metal phosphates (e.g., 2D $InPO_4$). While the focus of this study is on $P_2O_5$ intercalation, the possibility of intercalation from pre-dissociated molecules catalyzed by defects in graphene may exist for other types of molecules as well. This study is a significant milestone in advancing our understanding of intercalation routes of large molecules *via* the basal plane of graphene, as well as heterointerface chemical reactions leading to the formation of distinctive confined complex oxide compounds.

## INTRODUCTION

Intercalation is a topotactic insertion process of organic and/or inorganic chemical species (i.e., atoms, small molecules, etc.) between the interfaces and heterointerfaces of two-dimensional (2D) layered materials. Intercalation can occur through the exposed side edges of a 2D layered bulk crystal and/or through its basal planes[1]. In the latter case, for example, in monolayer to few-layer graphene, the intercalation pathways are typically point defects[2-8], and/or grain boundaries[9]. Usually the intercalation process can occur using a variety of processes, including vapor transport[10], wet-chemical[11-14], and electrochemical[15] means. To date, much of the research surrounding the intercalation of graphene has focused on the intercalation of metals (e.g., Ga, In, Al[16], Bi[17], Fe[18], Sb[19], Gd[20], Pb[21], etc.) and small gas molecules (e.g., $CO_2$[22], $O_2$[23]). However, the intercalation of molecules through the basal plane that are significantly larger than the lattice parameters of graphene remain highly unexplored. This is because the size and energy barrier for intercalation inhibits the permeation of such large molecules[24-25]. The ability to intercalate large molecules could further expand the toolsets for subsurface and heterointerface engineering of 2D layered materials, and therefore enrich the material choices available for the fabrication of heterostructures and intercalation compounds used in energy storage, optoelectronics, thermoelectric, catalysis, etc[26-28].

Chemical reactions enabled by confinement to the heterointerfaces of 2D layered materials provides a pathway to circumvent the limitation on intercalating large molecules directly through the basal plane. That is, individual chemical species or multiple small fragments of the initial state of the molecule would first intercalate through typically available pathways in the graphene lattice, and then recombine at the heterointerface *via* chemical reactions. So far, a variety of alloys ($Sn_{1-x}Ge_x$[29], $Fe_{1-x}Co_x$[30], etc.) and compounds ($GaN$[31-33], $AlN$[34], $MoS_2$[35], $PtSe_2$[36], $Ga_2O_3$[37], etc.) have been formed at graphene heterointerfaces *via* chemical reactions. In all these cases, however, chemical species are individually introduced in a sequential manner when intercalating graphene. This is done to prevent pre-reactions in the vapor phase between the individual elements prior to intercalation. Such pre-reactions usually lead to the formation of clusters at defect sites on the graphene surface, rather than intercalating through the lattice itself and forming a compound at the heterointerface.

In this work, we report a mechanism of vapor phase intercalation of large molecules through the heterointerface of monolayer graphene to form confined oxide materials underneath graphene with a chemical composition that resembles the initial state of the large intercalant molecule itself. Unlike prior work where chemical species and/or elements were sequentially introduced into the heterointerface, single molecules were introduced directly into the vapor phase for intercalation through the basal plane. This mechanism was revealed by investigating the intercalation of phosphorus pentoxide ($P_2O_5$) molecules directly from the vapor phase to form $P_2O_5$ at a graphene-substrate heterointerface. The formation of confined $P_2O_5$ was confirmed using a variety of chemical analysis and spectroscopic techniques as well as microscopic surface, and cross-sectional imaging. Density functional theory (DFT) were used to corroborate experimental results and to deduce the mechanism for the intercalation of $P_2O_5$ molecules through the basal plane of graphene. First, $P_2O_5$ dissociates into small fragments (i.e., $P_x$, $O_y$, etc.) when interacting with the graphene surface. These small fragments consistently showed lower energy barriers for intercalation through defects commonly observed in the graphene lattice. Once the small fragments intercalate, they react at the heterointerface to form $P_2O_5$. We show that the intercalation of $P_2O_5$ at the graphene heterointerface can tune the electronic structure of the graphene overlayer through strain and charge transfer. In addition, we show that this process can also be used in the formation of new confined metal phosphates by intercalating $P_2O_5$ into a graphene heterointerface initially containing confined 2D indium metal to form $InPO_4$. Although this study focuses on the intercalation of $P_2O_5$, the concept of intercalation from pre-dissociated molecules catalyzed by defects in graphene could be possible in other classes of molecules. Ultimately, this study provides an important steppingstone for advancing the understanding of intercalation pathways of large molecules though the basal plane of graphene as well as heterointerface chemical reactions to form unique confined complex oxide compounds.

**RESULTS AND DISCUSSION**

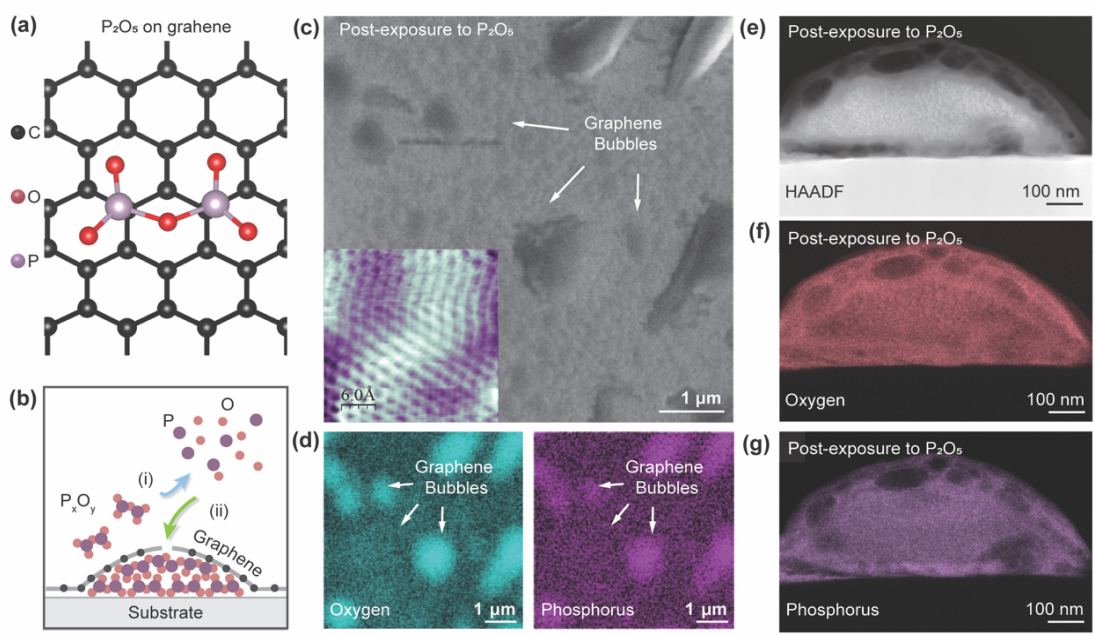

**Figure 1.** Intercalation of $P_2O_5$ at the heterointerface of graphene-germanium. (a) Configuration of $P_2O_5$ adsorbed onto pristine graphene calculated from DFT. (b) Schematic of intercalation process through the graphene heterointerface: (i) decomposition of $P_2O_5$ into $P_x$ and $O_y$, (ii) intercalation of $P_x$ and $O_y$ through graphene and formation of $P_2O_5$ at heterointerface *via* chemical reactions. (c) SEM highlighting the formation of graphene bubbles (white arrows), with STM image of the graphene lattice on the top surface of the bubble (bottom left inset). (d) SEM-EDS maps of (left) oxygen and (right) phosphorus of the highlighted graphene bubbles in (c). (e) STEM-HAADF cross-section image of a graphene bubble with corresponding STEM-EDS maps of the (f) Oxygen, and (g) Phosphorus distribution in the bubble.

The physical size of phosphorus pentoxide, a cage-like molecule with the empirical formula $P_2O_5$ (molecular formula $P_4O_{10}$), can be deduced from our DFT calculated ground state structure when adsorbed onto the surface of pristine graphene (**Figure 1a**). This molecule consists of two main phosphorus bonds with oxygen. The P-O single bond (denoted as oxygen bridging), has a bond length of $158\,pm$, and the P=O double bond (denotated as oxygen terminal) has a bond length of $143\ pm$ [38]. When intuitively comparing the physical size of the molecule to the lattice parameter of graphene, intercalation through the basal plane appears to be unlikely. Our DFT calculations revealed that the intercalation of $P_2O_5$ molecules in its native state through graphene could only occur if large pores, formed by the removal of 9 carbon atoms, were present (see **Supporting Information**). The formation of such large pores in graphene is, however, energetically unfavorable[39]. In fact, other theoretical studies have predicted large energy barriers for permeation of many atoms and molecules through pristine graphene[24-25]. In our case, we experimentally show that it is possible to intercalate $P_2O_5$ through monolayer graphene transferred onto germanium substrates without the need of large pores to serve as intercalation pathways. In our studies, germanium was used as a substrate due to the limited solubility[40] and low diffusivity[41] of phosphorous into germanium, allowing to fully capture and study the formation of $P_2O_5$ at graphene-germanium heterointerfaces. To intercalate $P_2O_5$, we prepared monolayer graphene on germanium substrates which were then annealed downstream of $SiP_2O_7$ at 950°C (see **Methods**). At such temperatures, the decomposition of $SiP_2O_7$ produced an upstream flux of $P_2O_5$ that impinges onto the surface of graphene downstream and subsequently intercalates into the heterointerface (**Figure 1b**). A more detailed discussion of the mechanism is included in the following sections.

As shown in **Figure 1c**, many bubbles appeared on the surface of graphene after exposing samples to a flux of $P_2O_5$ (*white arrows* in scanning electron micrograph, SEM). From scanning tunneling microscopy (STM) imaging of the bubble surface (**Figure 1c**, *inset*), the characteristic hexagonal lattice of graphene was clear. This suggests that not only was the graphene surface free of large pores, but also highlighted that the bubbles themselves were "graphene bubbles" containing subsurface material. Such graphene bubbles are prevalent in many intercalation studies of monolayer to few-layer graphene in the literature[24, 42]. Upon further inspection of the surface using energy dispersive spectroscopy (EDS), strong phosphorus and oxygen signals confined to the graphene bubbles were observed (**Figure 1d**). This implied the formation of phosphorus oxides within the bubbles themselves. The distribution of the phosphorus and oxygen in the graphene bubbles were further corroborated in sample cross-sections. In **Figure 1e**, a representative high-angle annular dark field scanning transmission electron microscope (HAADF-STEM) image of a graphene bubble is shown. The EDS cross-section elemental maps of the bubble (**Figure 1f and g**) also revealed strong oxygen and phosphorus signals. The atomic fractions of oxygen and phosphorus in the graphene bubble were 55.06% and 15.52%, respectively. This elemental distribution was further corroborated within x-ray photoelectron spectroscopy (XPS) depth profiles of the sample, facilitated by repeated ion sputtering (**Figure S2**). During the ion sputtering process, the graphene surface was first etched. This resulted in a rapid decrease in the C 1s core-level peak intensity (**Figure S2a**). The peak intensities for the Ge, O, and P core-levels increased upon sputtering of graphene (**Figure S2, b, c, and d**), which serves as additional evidence on the confinement of phosphorus oxide at the graphene-germanium substrate heterointerface after exposing samples to a flux of $P_2O_5$.

## Chemistry of the Graphene Heterointerface

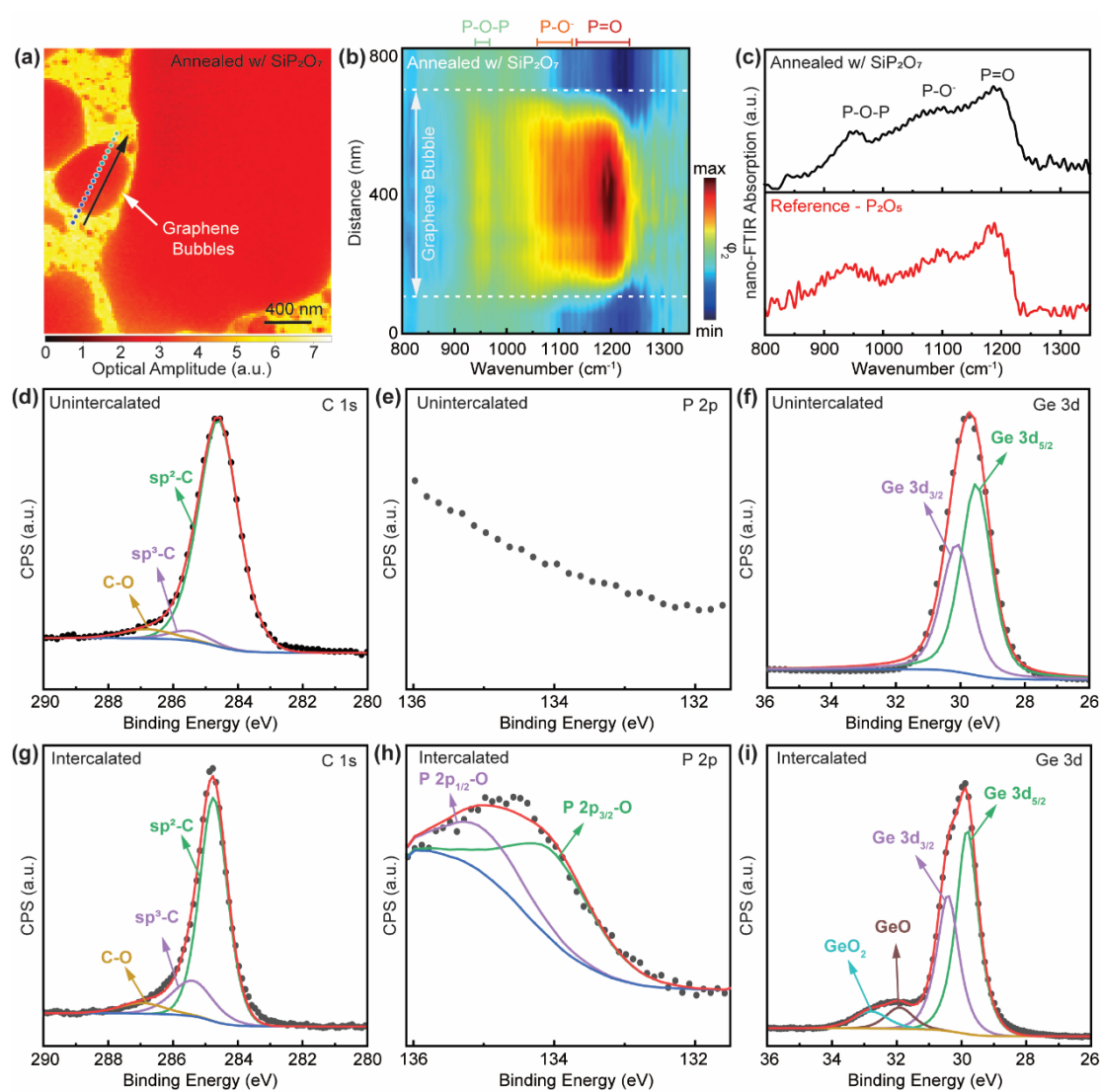

**Figure 2**. Chemistry at the graphene-germanium heterointerface. (a) Near-field white light IR image in a region of the sample containing large graphene bubbles. The white light IR image is the near-field amplitude ($S_2$) image, which is spectrally averaged over approximately ~610-1400 $cm^{-1}$ by setting the interferometer at the white light position. (b) Nano-FTIR spectral heat map acquired at 50 $nm$ intervals along the *black solid arrow* in (a). The graphene bubble region is highlighted with *white dashed lines* and *dashed arrow*. The collected phase channel ($\varphi_2$) was normalized to that obtained in a silicon sample using the same acquisition parameters. (c) Comparison between extracted nano-FTIR spectrum of a graphene bubble (top panel, *black solid line*) and spectrum collected from a reference P$_2$O$_5$ sample (bottom panel, *red solid line*). XPS of C, P and Ge core-levels from graphene on germanium samples (d, e, and f) annealed without an upstream flux of P$_2$O$_5$ (i.e., unintercalated sample) and from graphene on germanium samples (g, h, and i) annealed with an upstream flux of P$_2$O$_5$ (i.e., intercalated sample).

Moreover, the chemical environment of phosphorus oxide within the graphene bubbles was investigated using nano Fourier-transform infrared spectroscopy (nano-FTIR). A near-field white light IR image of a region of the sample containing graphene bubbles, spectrally averaged over $\sim 610 - 1400 \ cm^{-1}$, is highlighted in **Figure 2a**. Spectral points in the range associated with vibrational modes associated with the phosphorus and oxygen functional groups were acquired at 50 $nm$ intervals across a graphene bubble (*black arrow*). This is plotted in **Figure 2b** as a near-field phase ($\varphi_2$) heat map, which is related to the local absorption spectra of the material, and therefore, useful for determining the chemical environment of the intercalant[43-44]. **Figure 2c** highlights a comparison between two nano-FTIR spectra. The top panel (*black solid line*) is the extracted phase spectrum at the center of the graphene bubble in **Figure 2b**, while the bottom panel (*red solid line*) is a spectrum collected from a reference sample of $P_2O_5$ using high-resolution synchrotron-based nano-FTIR measurements (see **Methods**). Evident in **Figure 2b** and the top panel of **Figure 2c** are three characteristic IR bands localized to the graphene bubble. These bands were assigned as the P-O-P bending mode (940 - 970 $cm^{-1}$), the P-O bridging mode (1060 - 1125 $cm^{-1}$), and the P=O terminal oxygen stretching modes (1135 - 1235 $cm^{-1}$)[45]. The intensity of the P=O bond in the heat map (**Figure 2b**) reached a maximum at the center of the graphene bubble. More importantly, these peaks well matched the $P_2O_5$ reference sample in **Figure 2c** (bottom panel), and therefore supports phosphorus oxide that was confined at the graphene-germanium substrate heterointerface from the cross-section micrograph in **Figure 1g** was $P_2O_5$.

Furthermore, a comparative XPS study between non-intercalated and intercalated samples was also performed to further deduce the chemical bonding information at the graphene-germanium heterointerface. **Figure 2d-f** are XPS core-level spectra collected from samples that were annealed without an upstream flux of $P_2O_5$, while the XPS core-levels in **Figure 2g-i** were from samples annealed downstream to a flux of $P_2O_5$ (see **Methods**). In the non-intercalated samples, the $sp^3/sp^2$ carbon ratio, extracted from the C 1s core-level, was ~4% (**Figure 2d**), and neither phosphorus oxides (**Figure 2e**) nor germanium oxides (**Figure 2f**) could be detected in those samples. However, in intercalated samples, the $sp^3/sp^2$ carbon ratio increased to ~7.7% (**Figure 2g**). Such an increase in $sp^3$ bonded carbon within the non-intercalated samples was likely due to the formation of additional point defects in the graphene lattice. These defects could have resulted from the intercalation process itself and/or due to the enhanced hybridization of the graphene with the underlying intercalants. Moreover, a strong phosphorus oxide peak, whose binding energy was associated with $P_2O_5$ ($135\ eV$, **Figure 2h**), further corroborated its formation[46]. Also, from the Ge 3d core-level, the intercalated $P_2O_5$ appeared to oxidize the germanium substrate, resulting in characteristic peaks for GeO and $GeO_2$ that were observed between the binding energies of $31 - 33\ eV$ (**Figure 2i**). These combined results allude to a strong affinity of oxygen from $P_2O_5$ to the underlying germanium substrate, and perhaps provides an additional driving force in the intercalation process with the aid of point defect and/or grain boundaries in the graphene lattice as a pathway for intercalation through the basal plane of graphene[7].

**Mechanism of Large Molecule Intercalation Through Graphene**

To gain a deeper understanding into the mechanism by which phosphorus pentoxide $P_2O_5$ (or molecular formula $P_4O_{10}$) intercalates through the basal plane of graphene, we first examined the adsorptive properties of these molecules on five different graphene systems, consisting of pristine graphene and four graphene lattice defect configurations (**Figure 3a**, see **Supporting Information**). This survey was performed to factor in different interaction parameters of distinct chemical environments on the intercalation process. The configurational space of the adsorptive states (chemisorption *vs.* physisorption) was thoroughly explored by including a multitude of initial interacting geometries for both $P_2O_5$ and $P_4O_{10}$ on each of the different graphene lattice defect configurations (**Figure 3b**). As summarized in **Table S1**, covalent interactions with graphene occurred exclusively in the chemisorption of $P_2O_5$ onto graphene defect sites, whereas in pristine graphene, $P_2O_5$ interacted weakly *via* van der Waals (vdW) forces. In comparison, interactions between the more stable $P_4O_{10}$ molecule with all graphene systems were vdW in nature. In the case of covalent bonding between $P_2O_5$ adsorbents and graphene, P atoms consistently demonstrated a higher affinity towards graphene. This is shown by the greater adsorption energy of $P_2O_5$ to graphene through the P atom (denoted by $P_2O_5(P)$ in **Figure 3b**). In contrast, initial configurations of $P_2O_5$ approaching graphene with its O atom (designated by $P_2O_5(O)$) either relaxed to drastically different interaction geometries to allow bond formation between carbon and phosphorous or had a much lower total adsorption energy when compared to $P_2O_5(P)$. The only exception was $P_2O_5(O)$ chemisorbed to S-defects, in which both O and P atoms bonded covalently to the graphene defect site. Such trends can be rationalized by the fulfillment of covalency and increased charge transfer stabilization of carbon atoms at defects when bonding to less electronegative P atoms in $P_2O_5$.

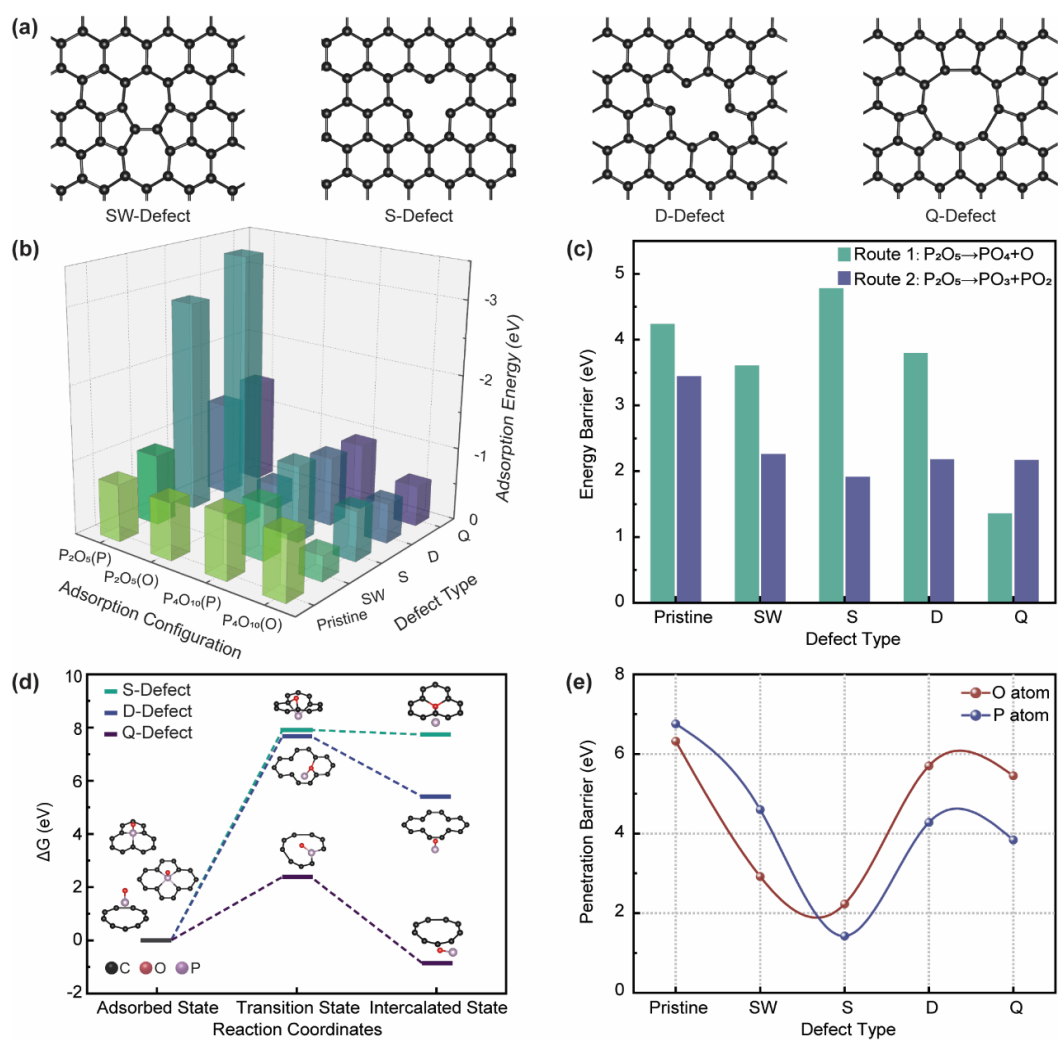

**Figure 3**. Mechanism for intercalation $P_2O_5$ through the basal plane of graphene (a) Top view of the four types of point defects in isolated graphene sheets. (b) Adsorption energy of phosphorus pentoxide on pristine and defective graphene monolayers. Within each $P_xO_y$

molecule, the atom interacting directly with carbon at the defect site is labeled in parentheses. (c) Calculated energy barrier of bond cleavage for P-O single bond (route 1, labelled as green) and double bond (route 2, labelled as blue) within $P_2O_5$ in different local environments. (d) Energy and configurational profile of PO molecule penetrating through defects. (e) Energy barrier of P (blue dots) and O (red dots) atoms permeating through defects.

Pristine graphene has been predicted to be impermeable to most gaseous species under non-extreme conditions[24-25, 42], and the presence of defects is thus the precondition for $P_2O_5$ intercalation. A quantitative description of intercalation through defects is complicated by the competition between diffusion and bonding with carbon atoms at defect sites. This gives rise to three different energy profiles corresponding to distinct pathways[47]. To address these complexities, we employed a case-by-case strategy by calculating the penetration barriers for fragments of $P_2O_5$ of varied size and chemistry based on calculated and experimental evidence from the literature[48]. Based on experimental evidence of $P_2O_5$ confinement at the graphene-germanium heterointerface, the intercalation of molecular $P_2O_5$ in its entirety was first investigated. Point defects of various sizes were examined, with the largest created by removing 9 adjacent carbon atoms in an isolated graphene sheet (**Figure S1**). In all scenarios, $P_2O_5$ decomposed into fragments solvated by carbon atoms at defect sites. Notably, the energetic profile of $P_2O_5$ permeating through large defects showed a negative reaction coordinate energy curve with small intermediate energy barriers as a result of the bonding of the fragmented $P_2O_5$ molecule onto the defect site. Despite this negative reaction coordinate energy curve, a barrier of at least $7.7\ eV$ was still present for fragments to desorb from the defect and diffuse into opposite sides of the monolayer graphene (i.e., intercalate). From this we ascertain that intercalation is only possible from significantly smaller molecular fragments of $P_2O_5$. Developed from these findings, we propose a mechanism involving three consecutive steps of realistic energy barriers to intercalate $P_2O_5$ through the basal plane of monolayer graphene. First, chemisorbed $P_2O_5$ decomposes into atoms or small molecular fragments. The resulting species then intercalate separately through point defects and then react underneath graphene monolayer to form $P_2O_5$ at the confined heterointerface.

Describing such processes with a full network of reaction pathways would require a dedicated theoretical effort and is beyond the scope of this study, but two modes of P-O bond cleavage representing the initial steps of two probable pathways were investigated to exemplify our hypothesis. As illustrated in **Figure 3c**, route 1 involves cleaving the P=O double bond to remove one terminal oxygen from the molecule, while route 2 describes the fragmentation of $P_2O_5$ into $PO_2$ and $PO_3$ by cleaving the single bond between P and bridging O. Energy barriers for both reactions are shown in **Figure 3c**. Evidently, with the exception of S-defect in route 1, all four types of point defects exhibited catalytic activities towards decomposition of $P_2O_5$. These findings were consistent with the results of adsorption state calculations, as covalent interactions and charge transfer between the defects and adsorbed molecules weaken the intramolecular bonds. Having substantiated the possibility of catalytic dissociation of adsorbed $P_2O_5$ at defect sites, we then calculated the energetics of smaller fragments intercalating through point defects. The simplest oxide of phosphorous, phosphorous monoxide (PO), was chosen to determine the lowest required barrier for any molecular species to permeate through the graphene defects (**Figure 3d**). Similar to $P_2O_5$, PO could not permeate through pristine graphene or Stone-Wales (SW) defects without introducing defects of larger sizes. When penetrating through S- and D-defects, the molecule decomposed into individual atoms, with the rate limiting step having an energy barrier as high as $8\ eV$. On the other hand, PO was surprisingly permeable at Q-defects. It diffused through Q-defect without further fragmentation, with a barrier of approximately $2\ eV$, comparable to the energy required to break down chemisorbed $P_2O_5$. Finally, to compare atomic phosphorous and oxygen diffusing through the four different types of defects in graphene (**Figure 3e**), we performed calculations on $(4 \times 6)$ graphene supercells and incorporated results for D- and Q-defects from the work by Song *et. al*[47]. Remarkably, while point defects enabled both species to intercalate with reduced energy barriers, the energy barrier for individual atoms to penetrate through D- and Q-defects were higher than those for the PO molecule fragments. This observation reiterates the complexity of the interaction process through graphene point defects that cannot be generalized into a single step process.

**The Impact of the Intercalants on the Physical Properties of Graphene**

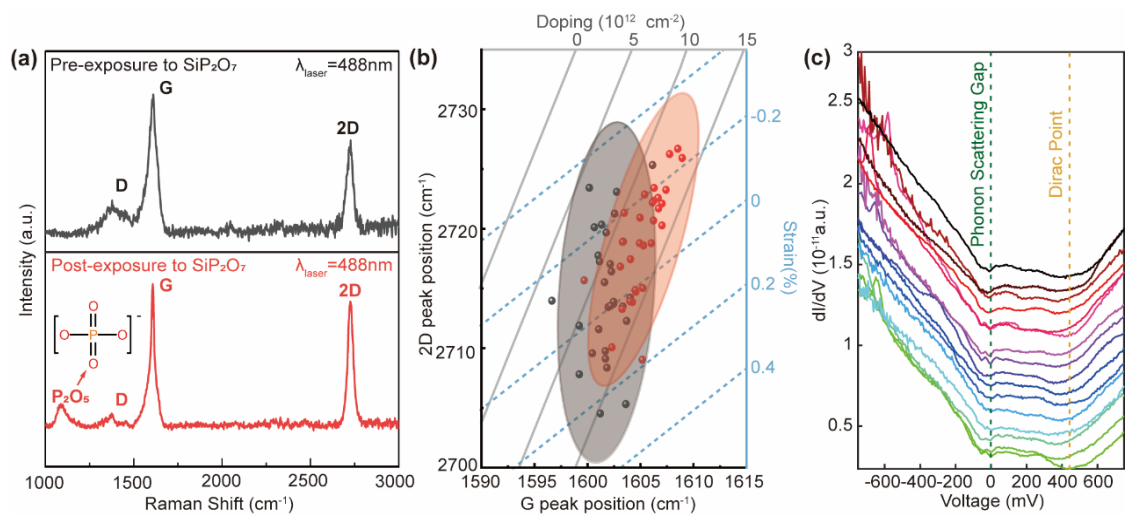

**Figure 4**. Raman spectra of graphene (a) Before (*black solid line*) and after (*red solid line*) exposure to an upstream flux of $P_2O_5$. (b) Raman 2D versus G peak position of graphene before (*black dots*) and after (*red dots*) exposure to an upstream flux of $P_2O_5$. (c) STS $dI/dV$ of intercalated graphene sample. The corresponding phonon scattering gap and Dirac points are highlighted with green dashed line and yellow dashed line, respectively. Spectra are vertically offset for clarity (Tunneling set point = 200 pA, 750 mV).

Graphene inherits highly delocalized π-electrons. Any modification to the spatial extent of the charge density of these π-electrons or tilt of the π-orbitals leads to significant changes to the physical properties of graphene[49-52]. For example, electronegative or electropositive intercalants confined underneath graphene will lead to charge transfer and doping. This in turn influences the spatial extent of the π-electron charge density and thus the Fermi level of graphene. Moreover, intercalants, such as those confined in graphene bubbles, can also lead to straining of the graphene lattice[53-55]. This will also change the π-orbital tilt and charge density in graphene and thus its Fermi level. Therefore, it is expected that the confinement of large $P_2O_5$ molecules at the graphene-germanium heterointerface would lead to significant changes to the physical properties of graphene itself. In our case, we show that charge transfer occurs from graphene to P₂O₅, resulting in p-type doping of the graphene layer. Raman spectroscopy was used to assess changes in the G and 2D peak positions that resulted from changes in strain and doping of graphene due to intercalation. In **Figure 4a**, we compare the Raman spectra between non-intercalated and intercalated samples. Besides the typical G, 2D and D peaks of graphene, a new Raman peak, approximately at 1100 $cm^{-1}$, was observed (**Figure 4a**, *red solid arrow*). This peak was associated with charge transfer from graphene to the intercalated P₂O₅, leading to the activation of $[PO_4]^-$ in the molecule[56]. Furthermore, it is possible to deconvolute the effects of charge transfer and strain on the doping level of graphene by performing a correlation analysis of the G and 2D peak positions between the non-intercalated and intercalated samples (**Figure 4b**). In the intercalated samples, a 12% increase in the hole doping level of graphene was observed when compared to the non-intercalated samples. This was further corroborated from scanning tunneling spectroscopy (STS) $dI/dV$ measurements of the intercalated graphene surface (**Figure 4c**). The $dI/dV$ curves were collected on the graphene surface in a line scan with a step of ~1.1 $nm$. The slope of the $dI/dV$ curve is proportional to the local density of states (LDOS) of the sample at the tunneling site. Two local minima were observed in all $dI/dV$ curves, that is, one at $V = 0$ which was the phonon scattering gap (*green dashed line*), and the another at the Dirac point where the LDOS of graphene reaches its minimum (*yellow dashed line*)[57]. From **Figure 4c**, the Dirac point was up-shifted to $+425\ mV$ (*yellow dashed line*) with respect to the Fermi level ($V = 0$) after intercalation, which indicated that P₂O₅ led to p-type doping of graphene.

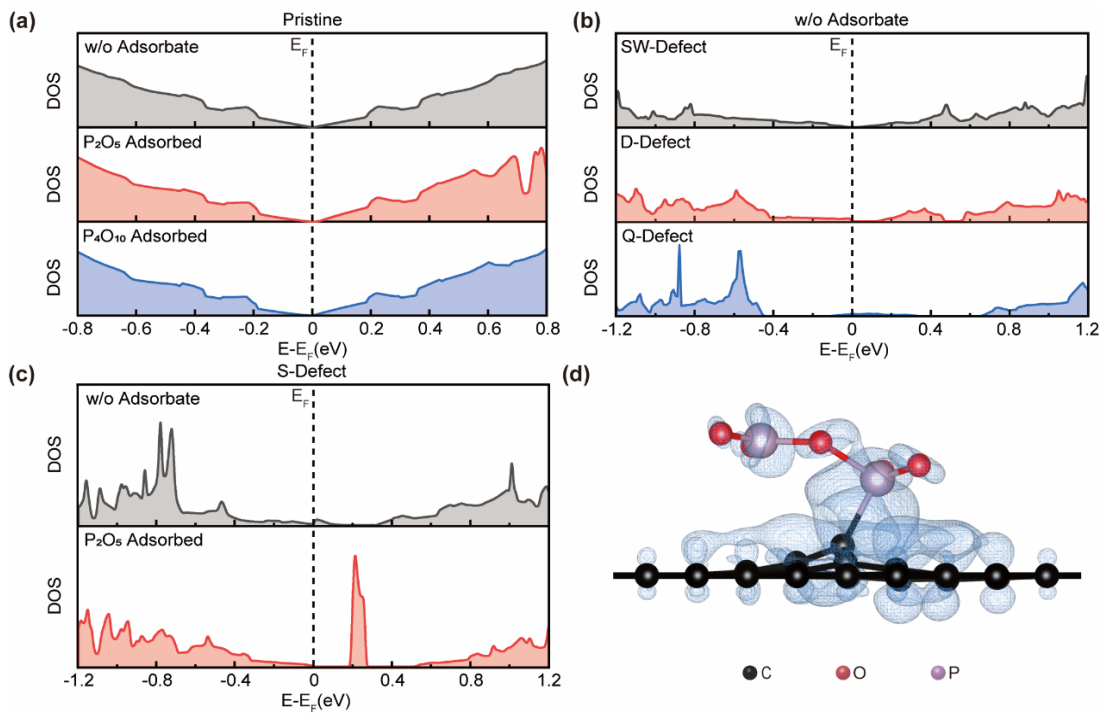

**Figure 5**. Projected DOS of (a) Pristine graphene sheet without (*black*) and with (*red and blue*) adsorbed phosphorous oxide, (b) Graphene with SW- (*black*), D- (*red*), and Q-defects (*blue*) without adsorbed phosphorous oxide. (c) Comparison of p-orbital DOS of graphene with S-defect (*black*) and with S-defect covalently bound to P atom in chemisorbed P₂O₅ (*red*). (d) Differential charge density of P₂O₅ covalently bound to S-defect. Blue wireframe isosurface denotes loss in electron charge equivalent to $0.001 e \cdot eV^{-1}$.

DFT calculations were also performed to investigate the origins of the observed conductive properties of graphene after intercalation. Given that graphene transferred onto germanium substrates could be subjected to strain[58-60], the impact of possible deformation on the electronic structure of monolayer graphene was studied. Our density of states (DOS) calculations based on (4 × 8) supercells of graphene suggested that the electronic structure near Fermi level did not undergo significant change, despite an increase in the overall system energy with respect to the increased strain. While compressive and tensile stress led to horizontal stretches and compression of the total DOS, respectively, the overall electronic structure near the Fermi level did not vary (**Figure S3**). We observed that both pristine and deformed graphene behaved as semimetals. As a result, strain alone did not significantly contribute to the experimentally observed p-type conductive properties in Raman and STS measurements. Similarly, while the exposure to $P_2O_5$ caused changes in DOS of graphene far above and below the Fermi level, the overall electronic properties of pristine graphene with adsorbed molecules remained semi-metallic, with characteristic converging overlap between the bottom of the conduction band and the peak of valence band (**Figure 5a**).

Moreover, to elucidate the effect of point defects on altering the bulk conductivity of graphene, the partial DOS of carbon atoms away from the defect sites were also calculated. As shown in **Figure 5b**, graphene with SW-defects had a conduction band overlapping the valence band at Fermi level, implying semi-metallic behaviors as similar to pristine graphene. On the contrary, systems with missing carbon atoms all developed bandgaps and mid-gap states near the Fermi level. Graphene with S-defects had a mid-gap state overlapping its conduction band, rendering it a semimetal, while graphene with D-defects behaved as a p-type semiconductor, where the Fermi level laid in the bandgap near conduction band. However, in the case of graphene with Q-defects, the Fermi level was well positioned in its mid-gap state, making it effectively a metal. Finally, we found that systems whose carbon atoms at defect sites formed covalent bonds to adsorbed $P_2O_5$ fragments underwent substantial changes in their electronic structure due to charge transfer between graphene and adsorbed molecules. DOS of carbon atoms in defected graphene bond to $P_2O_5$ is illustrated in **Figure S4**. In all defected systems, the presence of chemisorbed molecules served to enlarge the bandgap and enhance the mid-gap states, with $P_2O_5(P)$ chemisorbed to S-defect being the most notable case (**Figure 5c**). A summary of these effects is also listed in **Table S2**. To visualize the loss of electron density in the graphene π-orbitals when $P_2O_5$ was covalently bonded to the graphene defect sites, differential charge calculations were employed (**Figure 5d**). This change in electron charge density is equivalent to hole-doping of the graphene monolayer, which is consistent with our experimental observations.

**Synthesis of Metal Phosphates at Heterointerface**

Although phosphorus has limited solubility[40] and diffusivity[41] into germanium, our XPS analysis of the Ge 3d core-level revealed that the intercalation of $P_2O_5$ led to chemical reaction with the germanium substrate. This could have been possible from the intercalated oxygen fragments themselves and/or the affinity of $P_2O_5(O)$ to bond with germanium (**Figure 2i**). To further elucidate other chemical reactions of the intercalated fragments of $P_2O_5$ that could take place at graphene heterointerfaces, we show that this process can also be used to form confined metal phosphates by intercalating $P_2O_5$ into a graphene heterointerface initially containing 2D metals to form indium phosphate as the example in this study (see **Methods**). Indium phosphate, specifically $InPO_4$, is a wide bandgap insulator ($E_g = 4.5\ eV$)[61] with excellent dielectric properties that initially gain interest as a gate material for InP formed *via* surface oxidation[62-65]. However, the formation of dimensionally confined indium phosphates remains largely unexplored, thus impending a more comprehensive understanding of its physical properties and potential applications. In our case, we demonstrated the successful formation of confined indium phosphates by intercalating $P_2O_5$ into a graphene-SiC heterointerface initially containing confined mono- to bilayer indium also formed by intercalation[16, 66]. We identified the composition of the confined indium phosphate at heterointerface by investigating the In and P core-levels using high-resolution XPS (**Figure 6a** and **b**, respectively). In Figure 6a, we compare the core-levels of In 3d, before (*black solid line*) and after (*blue solid line*) intercalation of $P_2O_5$ to the graphene heterointerface containing 2D indium, to the In 3d core-level of a pure $In_2O_3$ reference sample (*red solid line*). In addition to the typical peaks for indium metal (*black dashed line*), the XPS spectra exhibited new peaks around $445.8\ eV$ and $453.3\ eV$ in the intercalated sample (*blue dashed line*). These two new peaks were up-shifted by $1.1\ eV$ from the peak position of the $In_2O_3$ reference (*red dashed line*)[67]. Furthermore, a strong phosphorus signal was detected in samples intercalated with $P_2O_5$. (**Figure 6b**) Therefore, based on these observations, we can exclude the formation $In_2O_3$ in the $P_2O_5$ intercalated samples. Further inspection and comparison of our XPS core-level spectra to previous reported XPS of indium phosphates suggest that the peaks at $445.8\ eV$ and $453.3\ eV$ are the In 3d core-levels associated with P-O-In $3d_{3/2}$ and P-O-In $3d_{5/2}$ bonding, respectively. Although the phase of the confined indium phosphate at the graphene heterointerface was likely in the form of $InPO_4$, $In(PO_3)_3$ and other metastable oxide phases could also form at such confined spaces[37, 68-69] and therefore of important investigation in future studies.

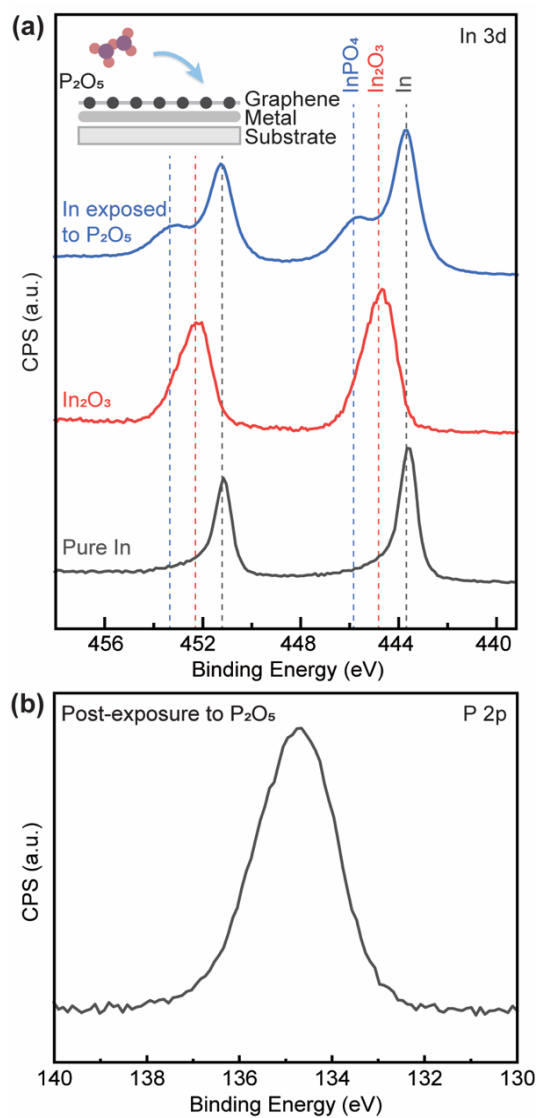

**Figure 6**. XPS core-levels of (a) In 3d for 2D indium confined at the graphene-SiC heterointerface before (*black solid line*) and after exposure to $P_2O_5$ (*blue solid line*). The XPS spectra for a pure $In_2O_3$ crystal (*red solid line*) is also included as a reference. (b) P 2p for the confined 2D indium after exposure to $P_2O_5$.

**CONCLUSION**

In summary, we demonstrated that large molecules, such as $P_2O_5$, can permeate the basal plane of graphene and intercalate into the heterointerface. This was evident from the strong P and O signals that were detected underneath the graphene surface from spectroscopic measurements (EDS, nano-FTIR, XPS, Raman, STS) and microscopic cross-sectional imaging and analysis in TEM. Although the physical size of the molecule was larger than that of the graphene lattice, fragments of the molecule could intercalate through typical pathways in the graphene lattice. This was revealed by calculating the penetration barriers for fragments of $P_2O_5$ of varied size and chemistry through pristine and defected graphene of different configurations. The complexity of the interaction process of large molecules through the basal plane of graphene cannot be generalized into a single step process. However, a two-step mechanism, based on realistic permeation energy barriers, was proposed for the intercalation of $P_2O_5$: (i) $P_2O_5$ first dissociated into small fragments (i.e., P, O, etc.) catalyzed by point defects in the graphene lattice; (ii) these fragments once intercalated can chemically react at the confined graphene-substrate heterointerface forming a condensed phase that chemically resembles the initial state of the molecule. It has been shown that the intercalation of $P_2O_5$ at the graphene heterointerface could effectively tune the doping level of graphene *via* charge transfer. Moreover, the intercalated $P_2O_5$ (and/or its fragments) can also act as reactants and further contribute to other interfacial reactions, for example, the conversion of 2D indium into 2D indium phosphate at the heterointerface between graphene and SiC. Although our findings only focus on indium phosphates, our results demonstrate the potential of this approach for the conversion of other metals (e.g., Ga, Al, etc.) and alloys. While the focus of this study is on $P_2O_5$ intercalation, the possibility of intercalation from pre-dissociated molecules catalyzed by defects in graphene may exist for other types of molecules as well. This study serves as a significant milestone in advancing the comprehension of intercalation routes of large molecules *via* the basal plane of graphene, as well as heterointerface chemical reactions leading to the formation of distinctive confined complex oxide compounds.

## METHODS

### Intercalation of $P_2O_5$ at graphene-substrate heterointerfaces:

The intercalation process was performed at 1 atm in a single zone furnace under argon flow. Two kinds of samples were used in this study: Transferred CVD graphene (Graphenea) onto cleaned Ge(110) substrates (MTI Corp.), and 2D indium samples intercalated at graphene-SiC(0001) heterointerface, as reported by Rajabpour et al.[16]. $SiP_2O_7$ (Saint-Gobain Inc.) was utilized to produce an upstream flux of $P_2O_5$ to the samples. First the center of the furnace was set to $950\,°C$ at a ramping rate of $55\,°C/min$ (stage-I) and then dwelled at $950\,°C$ for $45\,mins$ (stage-II). In stage-II, the measured temperature for the sample and $SiP_2O_7$ was $600\,°C$ and $950\,°C$, respectively. Samples were then naturally cooled down to room temperature under argon flow.

### Materials Characterization:

SEM and EDS were performed on a Thermo Fisher Scios 2 dual-beam microscope. The cross-section samples of graphene bubbles for TEM analysis were prepared using a focused ion beam (FIB) Zeiss NVision 40 FIB-SEM. The target location was extracted from the specimen and attached to a copper FIB grid using a conventional FIB milling and lift-out procedure. The HAADF image and STEM-EDS mappings on the cross-section of graphene bubbles were collected with a Thermo Fisher Scientific Talos $200X$ operated at $200\,kV$. HAADF was performed with a spot size less than $1\,nm$ with a convergence semi-angle of $10.5\,mrad$. Furthermore, XPS spectra and depth profiles were carried out in a Thermo-Fisher K-Alpha Plus XPS system and in Kratos Axis Ultra DLD system using a monochromatic Al $K\alpha$ source ($hv = 1486.6\,eV$). Raman measurements were performed on a Renishaw inVia system with a $488\,nm$ laser (maximum power at $100\,mW$). STM was performed in a closed-cycle RHK Pan Scan Freedom SPM system in UHV ($\sim 1 \times 10^{-10}\,mbar$) at a base temperature of $10\,K$. $dI/dV$ measurements were taken using a built-in digital lock-in amplifier operating at $1.3156\,kHz$ with a $10\,mV$ excitation amplitude. Moreover, nano-FTIR of the graphene bubbles was performed on a commercial scattering-type scanning near-field optical microscope (s-SNOM) setup (Neaspec Inc.) and the collected spectra were normalized to that obtained on a silicon sample using the same acquisition parameters. Nano-FTIR of the $P_2O_5$ reference sample was performed on a synchrotron infrared light-based nano-spectroscopy (SINS) setup (Innova, Bruker) (beamline 5.4, Advanced Light Source, Lawrence Berkeley National Laboratory).

### First Principles Calculation

DFT calculations were performed using generalized gradient approximation (GGA) with Vienna Ab initio Simulation Package. The Projector-Augmented-Wave (PAW) method was used and Grimme's DFT+D2 was applied to better account for the Van der Waal's interaction between graphene monolayer and absorbed species[70]. Kinetic energy cutoff was set at $500\,eV$, and the global convergence criterion for breaking electronic SC-loop was chosen to be $5 \times 10^{-5}$. A vacuum space of $22\,Å$ was applied on top of the graphene sheet to best model the isolated state of a monolayer. $\Gamma$ point sampling was used throughout all calculations. For relaxations and nudged elastic band (NEB) calculations, a grid of $5 \times 4 \times 1$ was used. To obtain density of states (DOS), a grid of 9 × 9 × 1 was adopted for best precision. For more information on simulating the reaction mechanism, see **Supporting Information**.

**ASSOCIATED CONTENT**

**Supporting Information**

Additional information including details of graphene transfer process and first principles calculation, intercalation of $P_2O_5$ molecule through a defect of 9 vacant atoms, XPS depth profile for graphene/Ge (110) after exposure to $P_2O_5$, projected DOS of carbon atoms in deformed graphene sheet, projected DOS of carbon in defected graphene with $P_2O_5$ adsorbed to atoms at defect site.

**AUTHOR INFORMATION**

**Corresponding Author**

**Zakaria Y. Al Balushi -** Department of Materials Science and Engineering, University of California, Berkeley, Berkeley, CA 94720, USA; Materials Sciences Division, Lawrence Berkeley National Laboratory, Berkeley, CA 94720, USA; Email: albalushi@berkeley.edu.

**Author**

**Jiayun Liang -** Department of Materials Science and Engineering, University of California, Berkeley, Berkeley, CA 94720, USA

https://orcid.org/0000-0002-4570-1017

**Ke Ma** - Department of Materials Science and Engineering, University of California, Berkeley, Berkeley, CA 94720, USA

https://orcid.org/0009-0001-2163-2686

**Xiao Zhao** - Department of Materials Science and Engineering, University of California, Berkeley, Berkeley, CA 94720, USA; Materials Sciences Division, Lawrence Berkeley National Laboratory, Berkeley, CA 94720, USA

https://orcid.org/0000-0003-1079-664X

**Guanyu Lu -** Department of Mechanical Engineering, Vanderbilt University, Nashville, TN 37235, USA

https://orcid.org/0000-0001-8960-0464

**Jake V. Riffle -** Department of Physics and Astronomy, University of New Hampshire, Durham, NH 03824, USA

https://orcid.org/0009-0007-2714-0133

**Carmen M. Andrei** - Canadian Centre for Electron Microscopy, McMaster University, Hamilton, ON L8S 4L8, Canada.

https://orcid.org/0000-0002-3093-8089

**Chengye Dong** - 2D Crystal Consortium, The Pennsylvania State University, University Park, PA 16802, USA

https://orcid.org/ 0000-0001-8598-5758

**Turker Furkan** - Department of Materials Science and Engineering, The Pennsylvania State University, University Park, PA 16802, USA

https://orcid.org/0000-0002-2385-8849

**Siavash Rajabpour** - Department of Materials Science and Engineering, The Pennsylvania State University, University Park, PA 16802, USA

https://orcid.org/0000-0002-1686-065X

**Rajiv Ramanujam Prabhakar**- Chemical Sciences Division, Lawrence Berkeley National Laboratory, Berkeley, CA 94720, USA

https://orcid.org/0000-0002-4598-9073

**Joshua A. Robinson** - Department of Materials Science and Engineering, The Pennsylvania State University, University Park, PA 16802, USA; 2D Crystal Consortium, The Pennsylvania State University, University Park, PA 16802, USA.

https://orcid.org/0000-0002-1513-7187

**Magdaleno R. Vasquez Jr. –** Department of Mining, Metallurgy, and Materials Engineering, University of the Philippines, Diliman, Quezon City 1101, Philippines

https://orcid.org/0000-0002-2208-428X

**Quang Thang Trinh –** Queensland Micro- and Nanotechnology Centre, Griffith University, Brisbane, 4111 Australia

https://orcid.org/0000-0002-3311-4691

**Joel W. Ager** - Materials Sciences Division, Lawrence Berkeley National Laboratory, Berkeley, CA 94720, USA; Department of Materials Science and Engineering, University of California, Berkeley, Berkeley, CA 94720, USA

https://orcid.org/0000-0001-9334-9751

**Miquel Salmeron** – Department of Materials Science and Engineering, University of California, Berkeley, Berkeley, CA 94720, USA; Materials Sciences Division, Lawrence Berkeley National Laboratory, Berkeley, CA 94720, USA

https://orcid.org/0000-0002-2887-8128

**Shaul Aloni** – The Molecular Foundry, Lawrence Berkeley National Laboratory, Berkeley, CA 94720, USA

https://orcid.org/0000-0002-7561-4336

**Joshua D. Caldwell –** Department of Mechanical Engineering, Vanderbilt University, Nashville, TN 37235, USA

https://orcid.org/0000-0003-0374-2168

**Shawna M. Hollen –** Department of Physics and Astronomy, University of New Hampshire, Durham, NH 03824, USA

https://orcid.org/0000-0002-9158-7876

**Hans A. Bechtel** – Advanced Light Source, Lawrence Berkeley National Laboratory, Berkeley, CA 94720, USA

https://orcid.org/0000-0002-7606-9333

**Nabil Bassim** – Canadian Centre for Electron Microscopy, McMaster University, Hamilton, ON L8S 4L8, Canada; Department of Materials Science and Engineering, McMaster University, Hamilton, ON L8S 4L8, Canada

https://orcid.org/0000-0002-9161-5769

**Matthew P. Sherburne** – Department of Materials Science and Engineering, University of California, Berkeley, Berkeley, CA 94720, USA

https://orcid.org/0000-0002-3992-1822

**Author Contributions**

J.L. performed all the intercalation experiments, SEM, SEM-EDS, and Raman characterization. K.M. and M.S. performed and analyzed all the theoretical calculations. C.D., T.F. and S.R. grew the epitaxial graphene and intercalated the indium under the supervision of J.A.R. C.A. and N.B. collected and analyzed all the cross-section STEM-EDS and HAADF results. G.L., J.C., H.B. and J.L. performed and analyzed the nano-FTIR characterization. X.Z., M.S., J.L., R.P., J.A. and S.A. performed and analyzed XPS characterization. J.R. and S.H. performed and analyzed STM and STS measurements. Z.A. conceived the idea and supervised the project. J.L., M.M. and Z.A. wrote the manuscript. All authors discussed, revised, and approved the manuscript.

**Notes**

The authors declare no competing financial interest.

**ACKNOWLEDGEMENT**

Z.Y.A. acknowledges the support of this work through the Laboratory Directed Research and Development (LDRD) Program of Lawrence Berkeley National Laboratory under U.S. Department of Energy Contract No. DE-AC02-05CH11231 and the Canadian Institute for Advanced Research (CIFAR) under the Azrieli Global Scholars Program in Quantum Materials. Work at the Molecular Foundry was supported by the Office of Science, Office of Basic Energy Sciences, of the U.S. Department of Energy under Contract No. DE-AC02-05CH11231. Some of the XPS analysis was performed under a collaboration with the Liquid Sunlight Alliance, which is supported by the U.S. Department of Energy, Office of Science, Office of Basic Energy Sciences, Fuels from Sunlight Hub under award number DESC0021266. This research used resources of the Advanced Light Source, which is a DOE Office of Science User Facility under no. DE-AC02-05CH11231. J. D. C. gratefully acknowledges support for this work from the Office of Naval Research grant N00014-22-12035. G. L. is supported through Army Research Office Small Business Technology Transfer (W911NF-22-P-0029). J.A.R, C.D., T.F. and S.R. acknowledge the support of the Air Force Office of Scientific Research (AFOSR) through contract FA9550-19-1-0295 and the 2D Crystal Consortium a National Science Foundation Materials Innovation Platform, under cooperative agreement DMR-1539916. M.P.S. acknowledges the support of the National Supercomputing Centre (NSCC) Singapore *via* the Project ID 13002533 and the support of the nanoQuench project (CHED PCARI IIID-2016-007).

Finally, the authors would thank Dr. Hao Chen for providing $In_2O_3$ XPS calibration data for this study.

**DATA AVAILABILITY STATEMENT**

All data needed to evaluate the conclusions in the paper are present in the manuscript and/or the Supporting Information.

**REFERENCES**

1. Whittingham, M. S. Intercalation Chemistry: An Introduction. *Intercalation chemistry* **1982,** 1-18.
2. Sutter, P.; Sadowski, J. T.; Sutter, E. A. Chemistry under Cover: Tuning Metal−Graphene Interaction by Reactive Intercalation. *Journal of the American Chemical Society* **2010,** *132*, 8175-8179.
3. Jin, L.; Fu, Q.; Mu, R.; Tan, D.; Bao, X. Pb Intercalation Underneath a Graphene Layer on Ru (0001) and Its Effect on Graphene Oxidation. *Physical Chemistry Chemical Physics* **2011,** *13*, 16655-16660.
4. Cui, Y.; Gao, J.; Jin, L.; Zhao, J.; Tan, D.; Fu, Q.; Bao, X. An Exchange Intercalation Mechanism for the Formation of a Two-Dimensional Si Structure Underneath Graphene. *Nano Research* **2012,** *5*, 352-360.
5. Jin, L.; Fu, Q.; Yang, Y.; Bao, X. A Comparative Study of Intercalation Mechanism at Graphene/Ru (0001) Interface. *Surface science* **2013,** *617*, 81-86.
6. Petrović, M.; Šrut Rakić, I.; Runte, S.; Busse, C.; Sadowski, J.; Lazić, P.; Pletikosić, I.; Pan, Z.-H.; Milun, M.; Pervan, P. The Mechanism of Caesium Intercalation of Graphene. *Nature communications* **2013,** *4*, 1-8.
7. Li, G.; Zhou, H.; Pan, L.; Zhang, Y.; Huang, L.; Xu, W.; Du, S.; Ouyang, M.; Ferrari, A. C.; Gao, H.-J. Role of Cooperative Interactions in the Intercalation of Heteroatoms between Graphene and a Metal Substrate. *Journal of the American Chemical Society* **2015,** *137*, 7099-7103.
8. Hui, J.; Burgess, M.; Zhang, J.; Rodríguez-López, J. Layer Number Dependence of $Li^+$ Intercalation on Few-Layer Graphene and Electrochemical Imaging of Its Solid–Electrolyte Interphase Evolution. *ACS nano* **2016,** *10*, 4248-4257.
9. Xia, C.; Watcharinyanon, S.; Zakharov, A.; Yakimova, R.; Hultman, L.; Johansson, L. I.; Virojanadara, C. Si Intercalation/Deintercalation of Graphene on 6H-SiC (0001). *Physical Review B* **2012,** *85*, 045418.
10. Bointon, T. H.; Khrapach, I.; Yakimova, R.; Shytov, A. V.; Craciun, M. F.; Russo, S. Approaching Magnetic Ordering in Graphene Materials by $FeCl_3$ Intercalation. *Nano letters* **2014,** *14*, 1751-1755.
11. Eda, G.; Fujita, T.; Yamaguchi, H.; Voiry, D.; Chen, M.; Chhowalla, M. Coherent Atomic and Electronic Heterostructures of Single-Layer $MoS_2$. *ACS nano* **2012,** *6*, 7311-7317.
12. Koski, K. J.; Wessells, C. D.; Reed, B. W.; Cha, J. J.; Kong, D.; Cui, Y. Chemical Intercalation of Zerovalent Metals into 2d Layered $Bi_2Se_3$ Nanoribbons. *Journal of the American Chemical Society* **2012,** *134*, 13773-13779.
13. Kovtyukhova, N. I.; Wang, Y.; Lv, R.; Terrones, M.; Crespi, V. H.; Mallouk, T. E. Reversible Intercalation of Hexagonal Boron Nitride with Brønsted Acids. *Journal of the American Chemical Society* **2013,** *135*, 8372-8381.
14. Voiry, D.; Salehi, M.; Silva, R.; Fujita, T.; Chen, M.; Asefa, T.; Shenoy, V. B.;

Eda, G.; Chhowalla, M. Conducting MoS$_2$ Nanosheets as Catalysts for Hydrogen Evolution Reaction. *Nano letters* **2013,** *13,* 6222-6227.
15. Ohzuku, T.; Iwakoshi, Y.; Sawai, K. Formation of Lithium‐Graphite Intercalation Compounds in Nonaqueous Electrolytes and Their Application as a Negative Electrode for a Lithium Ion (Shuttlecock) Cell. *Journal of The Electrochemical Society* **1993,** *140,* 2490.
16. Rajabpour, S.; Vera, A.; He, W.; Katz, B. N.; Koch, R. J.; Lassaunière, M.; Chen, X.; Li, C.; Nisi, K.; El‐Sherif, H. Tunable 2d Group-III Metal Alloys. *Advanced Materials* **2021,** *33,* 2104265.
17. Warmuth, J.; Bruix, A.; Michiardi, M.; Hänke, T.; Bianchi, M.; Wiebe, J.; Wiesendanger, R.; Hammer, B.; Hofmann, P.; Khajetoorians, A. A. Band-Gap Engineering by Bi Intercalation of Graphene on Ir (111). *Physical Review B* **2016,** *93,* 165437.
18. Shen, K.; Sun, H.; Hu, J.; Hu, J.; Liang, Z.; Li, H.; Zhu, Z.; Huang, Y.; Kong, L.; Wang, Y. Fabricating Quasi-Free-Standing Graphene on a SiC (0001) Surface by Steerable Intercalation of Iron. *The Journal of Physical Chemistry C* **2018,** *122,* 21484-21492.
19. Wolff, S.; Roscher, S.; Timmermann, F.; Daniel, M. V.; Speck, F.; Wanke, M.; Albrecht, M.; Seyller, T. Quasi‐Freestanding Graphene on SiC (0001) by Ar‐Mediated Intercalation of Antimony: A Route toward Intercalation of High‐Vapor‐Pressure Elements. *Annalen der Physik* **2019,** *531,* 1900199.
20. Link, S.; Forti, S.; Stöhr, A.; Küster, K.; Rösner, M.; Hirschmeier, D.; Chen, C.; Avila, J.; Asensio, M.; Zakharov, A. Introducing Strong Correlation Effects into Graphene by Gadolinium Intercalation. *Physical Review B* **2019,** *100,* 121407.
21. Klimovskikh, I. I.; Otrokov, M. M.; Voroshnin, V. Y.; Sostina, D.; Petaccia, L.; Di Santo, G.; Thakur, S.; Chulkov, E. V.; Shikin, A. M. Spin–Orbit Coupling Induced Gap in Graphene on Pt (111) with Intercalated Pb Monolayer. *ACS nano* **2017,** *11,* 368-374.
22. Granas, E.; Andersen, M.; Arman, M. A.; Gerber, T.; Hammer, B. r.; Schnadt, J.; Andersen, J. N.; Michely, T.; Knudsen, J. CO Intercalation of Graphene on Ir (111) in the Millibar Regime. *The Journal of Physical Chemistry C* **2013,** *117,* 16438-16447.
23. Zhang, H.; Fu, Q.; Cui, Y.; Tan, D.; Bao, X. Growth Mechanism of Graphene on Ru (0001) and O$_2$ Adsorption on the Graphene/Ru (0001) Surface. *The Journal of Physical Chemistry C* **2009,** *113,* 8296-8301.
24. Leenaerts, O.; Partoens, B.; Peeters, F. Graphene: A Perfect Nanoballoon. *Applied Physics Letters* **2008,** *93,* 193107.
25. Sun, P.; Yang, Q.; Kuang, W.; Stebunov, Y.; Xiong, W.; Yu, J.; Nair, R. R.; Katsnelson, M.; Yuan, S.; Grigorieva, I. Limits on Gas Impermeability of Graphene. *Nature* **2020,** *579,* 229-232.
26. Zhou, J.; Lin, Z.; Ren, H.; Duan, X.; Shakir, I.; Huang, Y.; Duan, X. Layered Intercalation Materials. *Advanced Materials* **2021,** *33,* 2004557.
27. Wan, J.; Lacey, S. D.; Dai, J.; Bao, W.; Fuhrer, M. S.; Hu, L. Tuning Two-Dimensional Nanomaterials by Intercalation: Materials, Properties and Applications. *Chemical Society Reviews* **2016,** *45,* 6742-6765.
28. Rajapakse, M.; Karki, B.; Abu, U. O.; Pishgar, S.; Musa, M. R. K.; Riyadh, S. S.; Yu, M.; Sumanasekera, G.; Jasinski, J. B. Intercalation as a Versatile Tool for Fabrication, Property Tuning, and Phase Transitions in 2d Materials. *npj 2D Materials and Applications* **2021,** *5,* 30.
29. Kim, H.; Dugerjav, O.; Lkhagvasuren, A.; Seo, J. M. Doping Modulation of Quasi-Free-Standing Monolayer Graphene Formed on SiC (0001) through Sn$_{1-x}$Ge$_x$ Intercalation. *Carbon* **2019,** *144,* 549-556.
30. Avvisati, G.; Gargiani, P.; Lizzit, D.; Valvidares, M.; Lacovig, P.; Petrillo, C.; Sacchetti, F.; Betti, M. G. Strong Ferromagnetic Coupling and Tunable Easy Magnetization Directions of Fe$_x$Co$_{1-x}$ Layer (S) Intercalated under Graphene. *Applied Surface Science* **2020,** *527,* 146599.

31. Al Balushi, Z. Y.; Wang, K.; Ghosh, R. K.; Vilá, R. A.; Eichfeld, S. M.; Caldwell, J. D.; Qin, X.; Lin, Y.-C.; DeSario, P. A.; Stone, G. Two-Dimensional Gallium Nitride Realized Via Graphene Encapsulation. *Nature materials* **2016,** *15*, 1166-1171.
32. Feldberg, N.; Klymov, O.; Garro, N.; Cros, A.; Mollard, N.; Okuno, H.; Gruart, M.; Daudin, B. Spontaneous Intercalation of Ga and in Bilayers During Plasma-Assisted Molecular Beam Epitaxy Growth of GaN on Graphene on Sic. *Nanotechnology* **2019,** *30*, 375602.
33. Wang, W.; Li, Y.; Zheng, Y.; Li, X.; Huang, L.; Li, G. Lattice Structure and Bandgap Control of 2d GaN Grown on Graphene/Si Heterostructures. *Small* **2019,** *15*, 1802995.
34. Wang, W.; Zheng, Y.; Li, X.; Li, Y.; Zhao, H.; Huang, L.; Yang, Z.; Zhang, X.; Li, G. 2d Aln Layers Sandwiched between Graphene and Si Substrates. *Advanced Materials* **2019,** *31*, 1803448.
35. Michałowski, P. P.; Knyps, P.; Ciepielewski, P.; Caban, P. A.; Dumiszewska, E.; Kowalski, G.; Tokarczyk, M.; Baranowski, J. M. Growth of Highly Oriented $MoS_2$ Via an Intercalation Process in the Graphene/Sic (0001) System. *Physical Chemistry Chemical Physics* **2019,** *21*, 20641-20646.
36. Fu, J.; Hong, M.; Shi, J.; Xie, C.; Jiang, S.; Shang, Q.; Zhang, Q.; Shi, Y.; Huan, Y.; Zhang, Z. Intercalation-Mediated Synthesis and Interfacial Coupling Effect Exploration of Unconventional Graphene/$PtSe_2$ Vertical Heterostructures. *ACS applied materials & interfaces* **2019,** *11*, 48221-48229.
37. Turker, F.; Dong, C.; Wetherington, M. T.; El-Sherif, H.; Holoviak, S.; Trdinich, Z. J.; Lawson, E. T.; Krishnan, G.; Whittier, C.; Sinnott, S. B. 2d Oxides Realized Via Confinement Heteroepitaxy. *Advanced Functional Materials* **2023,** *33*, 2210404.
38. Hoppe, U.; Walter, G.; Barz, A.; Stachel, D.; Hannon, A. The PO Bond Lengths in Vitreous Probed by Neutron Diffraction with High Real-Space Resolution. *Journal of Physics: Condensed Matter* **1998,** *10*, 261.
39. Tian, W.; Li, W.; Yu, W.; Liu, X. A Review on Lattice Defects in Graphene: Types, Generation, Effects and Regulation. *Micromachines* **2017,** *8*, 163.
40. Olesinski, R.; Kanani, N.; Abbaschian, G. The Ge− P (Germanium-Phosphorus) System. *Bulletin of Alloy Phase Diagrams* **1985,** *6*, 262-266.
41. Cai, Y.; Camacho-Aguilera, R.; Bessette, J. T.; Kimerling, L. C.; Michel, J. High Phosphorous Doped Germanium: Dopant Diffusion and Modeling. *Journal of Applied Physics* **2012,** *112*, 034509.
42. Bunch, J. S.; Verbridge, S. S.; Alden, J. S.; Van Der Zande, A. M.; Parpia, J. M.; Craighead, H. G.; McEuen, P. L. Impermeable Atomic Membranes from Graphene Sheets. *Nano letters* **2008,** *8*, 2458-2462.
43. Huth, F.; Govyadinov, A.; Amarie, S.; Nuansing, W.; Keilmann, F.; Hillenbrand, R. Nano-Ftir Absorption Spectroscopy of Molecular Fingerprints at 20 nm Spatial Resolution. *Nano letters* **2012,** *12*, 3973-3978.
44. Bechtel, H. A.; Muller, E. A.; Olmon, R. L.; Martin, M. C.; Raschke, M. B. Ultrabroadband Infrared Nanospectroscopic Imaging. *Proceedings of the National Academy of Sciences* **2014,** *111*, 7191-7196.
45. Dayanand, C.; Bhikshamaiah, G.; Tyagaraju, V. J.; Salagram, M.; Krishna Murthy, A. Structural Investigations of Phosphate Glasses: A Detailed Infrared Study of the x(PbO)-(1−x)$P_2O_5$ Vitreous System. *Journal of materials science* **1996,** *31*, 1945-1967.
46. Wang, Y.; Sherwood, P. M. Phosphorus Pentoxide ($P_2O_5$) by XPS. *Surface Science Spectra* **2002,** *9*, 159-165.
47. Song, B.; Pan, L. Penetration of the First-Two-Row Elements through Mono-Layer Graphene. *Carbon* **2016,** *109*, 117-123.
48. Flemish, J.; Tressler, R. Pxoy Evaporation from $SiP_2O_7$ and Its Relationship to Phosphosilicate Glass Films. *Journal of the Electrochemical Society* **1991,** *138*, 3743.
49. Sreeprasad, T.; Berry, V. How Do the Electrical Properties of Graphene Change with Its Functionalization? *small* **2013,** *9*, 341-350.

50. Wu, Q.; Wu, Y.; Hao, Y.; Geng, J.; Charlton, M.; Chen, S.; Ren, Y.; Ji, H.; Li, H.; Boukhvalov, D. W. Selective Surface Functionalization at Regions of High Local Curvature in Graphene. *Chemical communications* **2012**, *49*, 677-679.
51. Huttmann, F.; Martínez-Galera, A. J.; Caciuc, V.; Atodiresei, N.; Schumacher, S.; Standop, S.; Hamada, I.; Wehling, T. O.; Blügel, S.; Michely, T. Tuning the Van Der Waals Interaction of Graphene with Molecules Via Doping. *Physical review letters* **2015**, *115*, 236101.
52. Banerjee, S.; Rappe, A. M. Mechanochemical Molecular Migration on Graphene. *Journal of the American Chemical Society* **2022**, *144*, 7181-7188.
53. Levy, N.; Burke, S.; Meaker, K.; Panlasigui, M.; Zettl, A.; Guinea, F.; Neto, A. C.; Crommie, M. F. Strain-Induced Pseudo-Magnetic Fields Greater Than 300 Tesla in Graphene Nanobubbles. *Science* **2010**, *329*, 544-547.
54. Khestanova, E.; Guinea, F.; Fumagalli, L.; Geim, A.; Grigorieva, I. Universal Shape and Pressure inside Bubbles Appearing in Van Der Waals Heterostructures. *Nature communications* **2016**, *7*, 1-10.
55. Sanchez, D. A.; Dai, Z.; Lu, N. 2d Material Bubbles: Fabrication, Characterization, and Applications. *Trends in Chemistry* **2021**, *3*, 204-217.
56. Hudgens, J. J.; Brow, R. K.; Tallant, D. R.; Martin, S. W. Raman Spectroscopy Study of the Structure of Lithium and Sodium Ultraphosphate Glasses. *Journal of non-crystalline solids* **1998**, *223*, 21-31.
57. Zhang, Y.; Brar, V. W.; Wang, F.; Girit, C.; Yayon, Y.; Panlasigui, M.; Zettl, A.; Crommie, M. F. Giant Phonon-Induced Conductance in Scanning Tunnelling Spectroscopy of Gate-Tunable Graphene. *Nature Physics* **2008**, *4*, 627-630.
58. Dai, J.; Wang, D.; Zhang, M.; Niu, T.; Li, A.; Ye, M.; Qiao, S.; Ding, G.; Xie, X.; Wang, Y. How Graphene Islands Are Unidirectionally Aligned on the Ge (110) Surface. *Nano Letters* **2016**, *16*, 3160-3165.
59. Chen, W.; Wang, X.; Li, S.; Yan, C.; He, L.; Zhang, P.; Yang, Y.; Ma, D.; Nie, J.; Dou, R. Robust Atomic-Structure of the 6× 2 Reconstruction Surface of Ge (110) Protected by the Electronically Transparent Graphene Monolayer. *Physical Chemistry Chemical Physics* **2020**, *22*, 22711-22718.
60. Sutter, P.; Hybertsen, M.; Sadowski, J.; Sutter, E. Electronic Structure of Few-Layer Epitaxial Graphene on Ru (0001). *Nano letters* **2009**, *9*, 2654-2660.
61. Wager, J.; Wilmsen, C.; Kazmerski, L. Estimation of the Band Gap of Inpo4. *Applied physics letters* **1983**, *42*, 589-591.
62. Torii, Y.; Hattori, K. On the Electrical Properties of the $InP_xO_y$-InP Interface. *Thin solid films* **1991**, *202*, 29-37.
63. Albanesi, E.; Sferco, S.; Lefebvre, I.; Allan, G.; Lannoo, M. Electronic Structure of Crystalline InP Oxides. *Solid state communications* **1993**, *86*, 27-31.
64. Simon, N.; Pascanut, C. D.; Santinacci, L.; Goncalves, A.-M.; Joudrier, A.-L.; Etcheberry, A. Morphology, Composition and Electrical Properties of Thin Anodic Oxides on InP. *ECS Transactions* **2009**, *19*, 273.
65. Ouerdane, A.; Bouslama, M.; Ghaffour, M.; Abdellaoui, A.; Hamaida, K.; Lounis, Z.; Monteil, Y.; Berrouachedi, N.; Ouhaibi, A. Study by EELS and EPES of the Stability of $InPO_4$/InP System. *Applied surface science* **2008**, *254*, 7394-7400.
66. Briggs, N.; Bersch, B.; Wang, Y.; Jiang, J.; Koch, R. J.; Nayir, N.; Wang, K.; Kolmer, M.; Ko, W.; De La Fuente Duran, A. Atomically Thin Half-Van Der Waals Metals Enabled by Confinement Heteroepitaxy. *Nature materials* **2020**, *19*, 637-643.
67. Chen, H.; Blatnik, M. A.; Ritterhoff, C. L.; Sokolović, I.; Mirabella, F.; Franceschi, G.; Riva, M.; Schmid, M.; Čechal, J.; Meyer, B. Water Structures Reveal Local Hydrophobicity on the $In_2O_3$ (111) Surface. *ACS nano* **2022**, *16*, 21163-21173.
68. Hollinger, G.; Bergignat, E.; Joseph, J.; Robach, Y. On the Nature of Oxides on InP Surfaces. *Journal of Vacuum Science & Technology A: Vacuum, Surfaces, and Films* **1985**, *3*, 2082-2088.
69. Faur, M.; Faur, M.; Jayne, D.; Goradia, M.; Goradia, C. XPS Investigation of

Anodic Oxides Grown on P‑Type InP. *Surface and Interface Analysis* **1990,** *15*, 641-650.

70. Grimme, S. Semiempirical GGA‑Type Density Functional Constructed with a Long‑Range Dispersion Correction. *Journal of computational chemistry* **2006,** *27*, 1787-1799.

## SUPPORTING TEXT

### Graphene Transfer Process

1cm by 1cm CVD-grown monolayer graphene (MLG) on polymer film (Graphenea) was placed into deionized water slowly and detached from the supporting polymer film underneath. Afterwards, a clean Ge (110) substrate was then introduced into deionized water and fished the graphene from below, followed by 30-mins dry in air and another 1-hour heating treatment at 150 °C on a hotplate. Later, graphene-Ge (110) sample was kept in low pressure environment (~$10^{-3}$ Torr) overnight and annealed at 450 °C in $N_2$ atmosphere for 2 hours.

### Details on First Principles Calculation

When simulating the dissociation of phosphorous oxide molecules and hetero atom intercalation through defected graphene monolayer, climbing image nudged elastic band (CI-NEB) calculations were performed to determine the most probable reaction route[1]. Such process involves three steps. First, the most stable initial and final configurations were obtained by structural relaxation. Then a CI-NEB calculation with 6-8 intermediates was conducted to find a most realistic energy path connecting the two configurations. Lastly, the two intermediates closest to the maxima on energy profile were used as initial and final configurations for another NEB calculation of 3-4 images to precisely determine the geometry and energy profile of intermediate states.

### Intrinsic Defects in Monolayer Graphene

Four types of intrinsic defects of graphene were studied via DFT. Stone-Wales defects (denoted as SW-Defects in **Fig. 4a**) are created by rotating a single pair of carbon atoms by 90 degrees to form adjacent pairs of pentagonal and heptagonal rings. The formation of Stone-Wales defects does not involve the introduction or removal of any atoms. Single vacancy defect (denoted as S-Defect in **Fig. 4a**) is formed when a single carbon atom is removed from a hexagon ring in graphene monolayer. In response to the removal of single atom, the structure near the vacancy undergoes Jahn-Teller distortion to minimize total energy, leading to a decreased separation between carbon atoms near the vacancy. Similarly, double vacancy defects (D-Defects in **Fig. 4a**) are formed through the loss of two adjacent carbon atoms in pristine monolayer graphene. Lastly, quadruple vacancy defects (Q-Defects in **Fig. 4a**) are produced by removing one carbon atom and its three nearest neighbors from the hexagon, resulting in a total of four missing atoms.

## SUPPORTING REFERENCES

1. Henkelman, G.; Uberuaga, B. P.; Jónsson, H. A Climbing Image Nudged Elastic Band Method for Finding Saddle Points and Minimum Energy Paths. *The Journal of chemical physics* **2000,** *113*, 9901-9904.

**SUPPORTING TABLES**

**Table S1.** Summary of absorption mechanisms of $P_2O_5$ and $P_4O_{10}$ on pristine and defected graphene monolayer. The notation (P) denotes that the molecule is binding or approaching graphene primarily through the phosphorous atom, while (O) represents adsorption primarily through the oxygen atom.

| Adsorption Mechanism | $P_2O_5$(P) | $P_2O_5$(O) | $P_4O_{10}$(P) | $P_4O_{10}$(O) |
|---|---|---|---|---|
| Pristine Graphene | vdW | vdW | vdW | vdW |
| SW-Defect | Covalent | Reaction | vdW | vdW |
| S-Defect | Covalent | Covalent | vdW | vdW |
| D-Defect | Covalent | vdW | vdW | vdW |
| Q-Defect | Covalent | Reaction | vdW | vdW |

**Table S2.** Conductive properties of pristine and defected graphene sheets exposed to $P_2O_5$ molecules.

| | Without Absorbate | $P_2O_5$(P) | $P_2O_5$(O) |
|---|---|---|---|
| Pristine Graphene | Semimetal | Semimetal | Semimetal |
| SW-Defect | Semimetal | Semimetal | Semimetal |
| S-Defect | Semimetal | P-type | P-type |
| D-Defect | P-type | Semimetal | P-type |
| Q-Defect | Metal | Metal | Metal |

**SUPPORTING FIGURES**

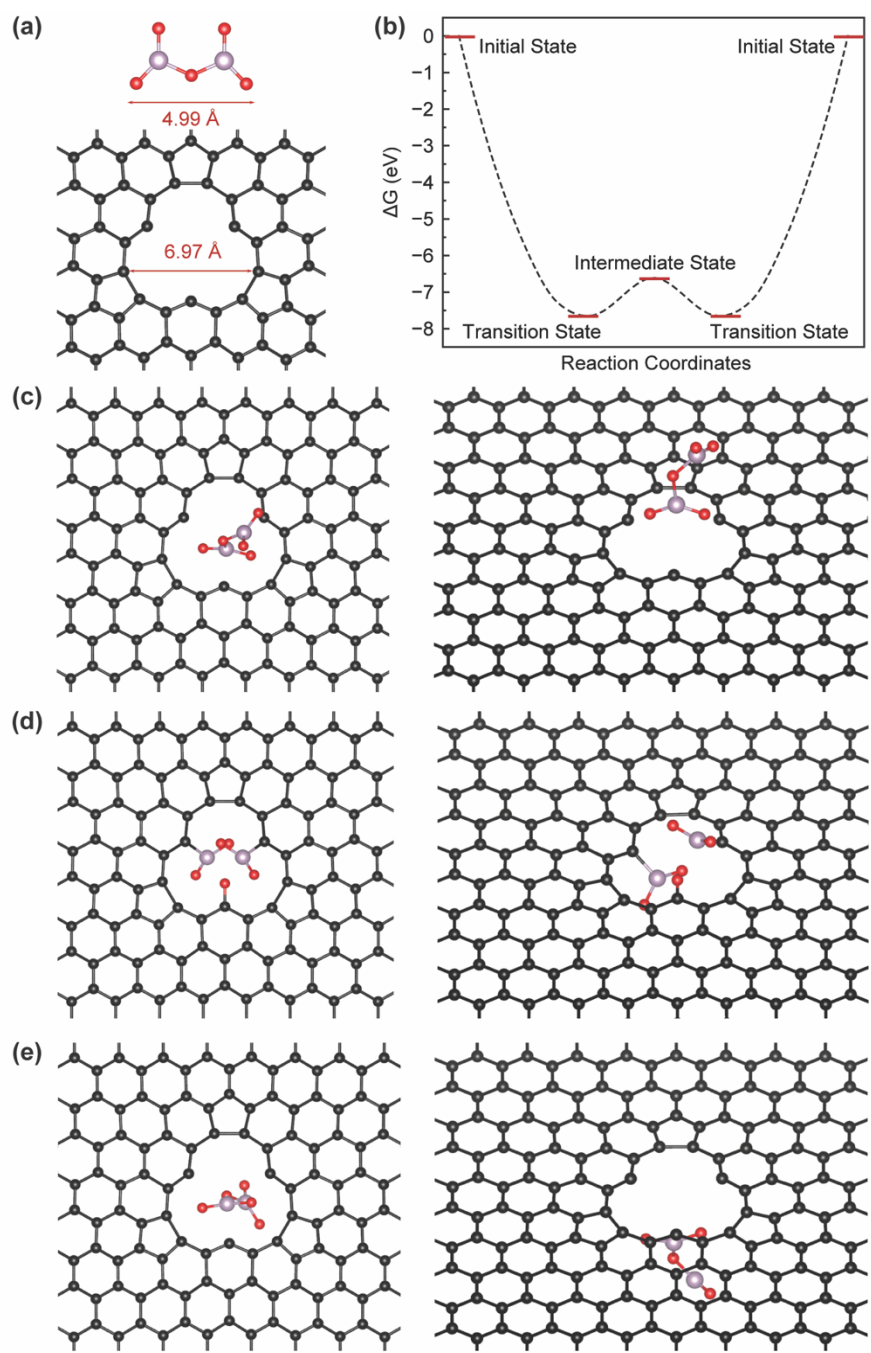

**Figure S4.** Intercalation of $P_2O_5$ molecule through a defect of 9 vacant atoms. (a) A comparison of dimensions of a defect of 9 vacant atoms and $P_2O_5$ molecule. (b) Energy profile of $P_2O_5$ intercalation through a 9-vacant-atom defect. The molecule undergoes spontaneous dissociation upon contact with the defect site and would require an energy barrier of 7.7eV to desorb and recombine on the opposite side of the graphene monolayer. Top (*left*) and angled (*right*) view of the (c) Initial, (d) Intermediate, and (e) Final state configuration of $P_2O_5$ molecule during intercalation.

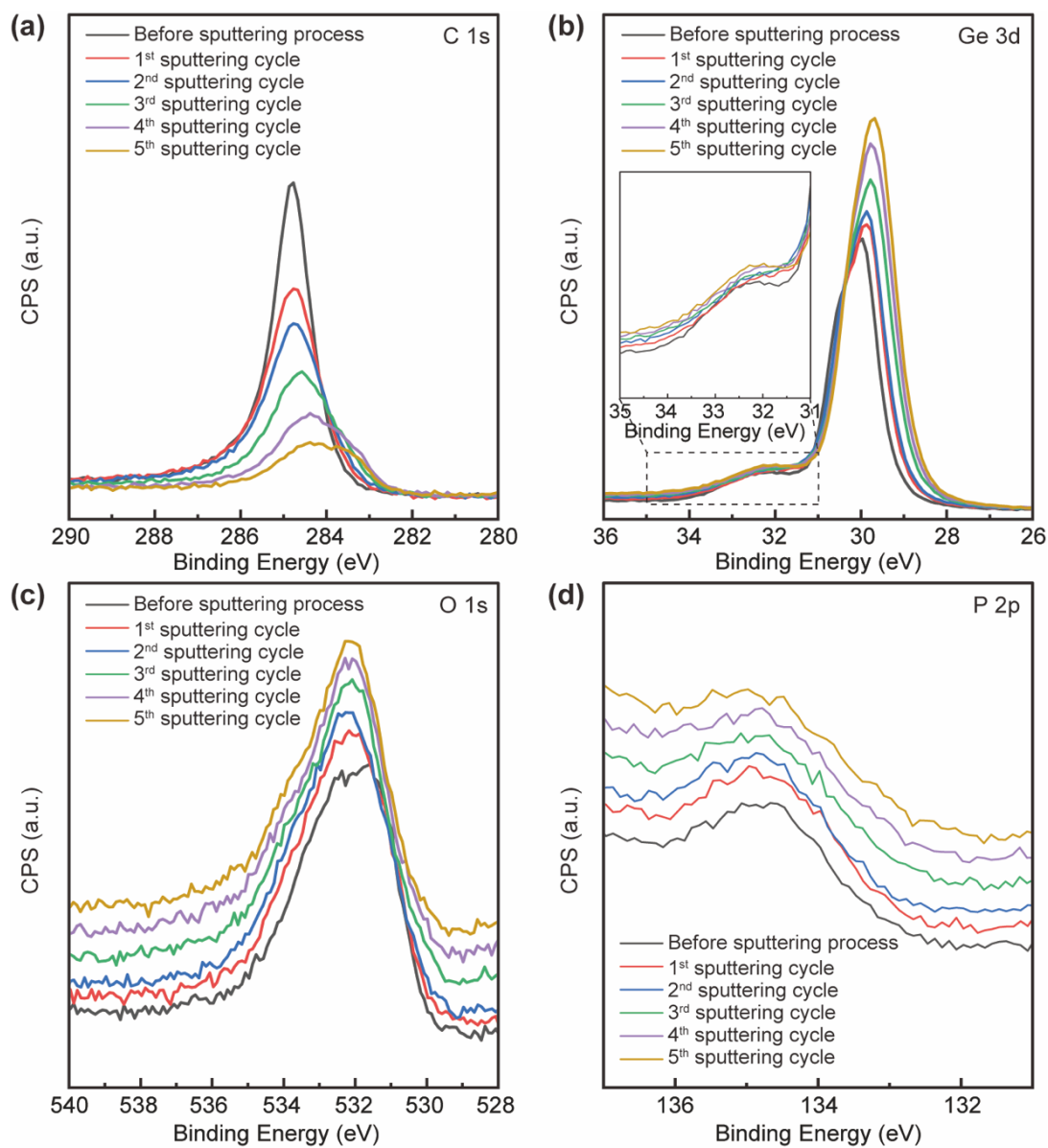

**Figure S5.** XPS depth profile of graphene-germanium after exposure to $P_2O_5$. High-resolution XPS spectra collected before the ion sputtering process (*black*); after the first sputtering cycle (*red*); after the second sputtering cycle (*blue*); after the third sputtering cycle (*green*); after the fourth sputtering cycle (*violet*) and after the fifth sputtering cycle (*yellow*) for core-levels of (a) C 1s, (b) Ge 3d, (c) O 1s, and (d) P 2p. In the XPS spectra of Ge 3d, region where the binding energy ranged from 31 eV to 35 eV was magnified to highlight the intensity change of the germanium oxide.

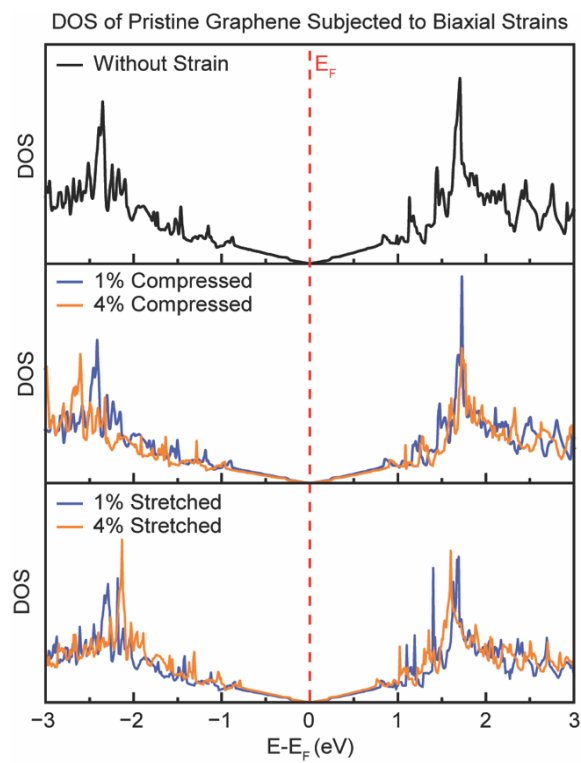

**Figure S6.** Projected DOS of carbon atoms in deformed graphene monolayer.

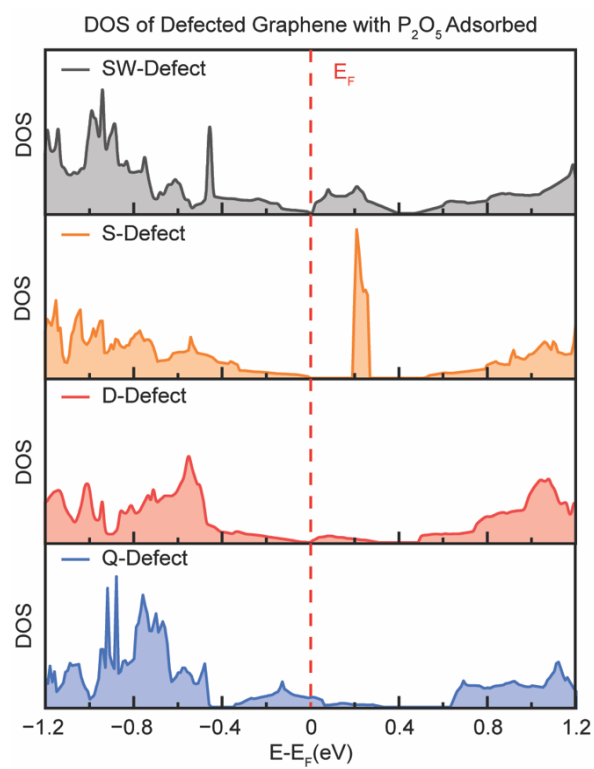

**Figure S7**. Projected DOS of carbon in defected graphene monolayer with $P_2O_5$ adsorbed to atoms at defect site.